\documentclass[aps,prb,twocolumn]{revtex4}
\usepackage{graphics,epsfig,amsfonts,amssymb,amsmath}
\begin{document}

\title{Non-resonant Raman response of inhomogeneous structures in the electron doped
$t-t'$ Hubbard model
 }
\author{B. Valenzuela}
\affiliation{Instituto de Ciencia de Materiales de Madrid,
CSIC. Cantoblanco. E-28049 Madrid, Spain}
\begin{abstract}
We calculate the non-resonant Raman response, the single particle
spectra and the charge-spin configuration for the electron doped
$t-t'$ Hubbard model using unrestricted Hartree-Fock calculations.
We discuss the similarities and differences in the response of
homogeneous versus inhomogeneous structures. Metallic
antiferromagnetism dominates in a large region of the $U-n$ phase
diagram but at high values of the on-site interaction and for
intermediate doping values inhomogeneous configurations are found
with lower energy. This result is in contrast with the case of hole
doped cuprates where inhomogeneities are found already at very low
doping. The inhomogeneities found are in-phase stripes compatible
with inelastic neutron scattering experiments. They give an
incoherent background in the Raman response. The $B_{2g}$ signal can
show a quasiparticle component even when no Fermi surface is found
in the nodal direction.
\end{abstract}

\date{\today}
\maketitle
\section{Introduction}
Electron and hole doped cuprates share a similar phase diagram with
antiferromagnetic and superconducting phases. However there are
quantitative differences between them. While electron-doped show
commensurate antiferromagnetism\cite{shirane} in a large region of
the $U-n$ phase diagram there is a quick suppression of
antiferromagnetism in hole-doped cuprates. There it is well
established that inhomogeneities appear when doping the Mott
insulator. Among them the stripes, where holes are aligned in a row,
have acquired importance as a possible scenario to explain the phase
diagram of the hole doped cuprates\cite{kivelson}. The strongest
experimental support for stripes comes from neutron scattering
experiments\cite{NScatTr} that show incommensurate peaks. They have
been interpreted as coming from anti-phase stripes where the line of
holes separate antiferromagnetic domains of opposite sign. The case
of the electron doped cuprates is less clear. The signal obtained in
neutron scattering experiments  is commensurate\cite{NScat} what
would be compatible with in-phase inhomogeneities. This possibility
can convene  with  theoretical predictions that incommensurate
solutions will be more favored in hole than in electron
doping\cite{tohmaek94,leung94,moreo,zhang,oles}. Very recent
measurements  of thermal conductivity and neutron
scattering\cite{ando} in electron doped cuprates have been
interpreted as in-phase stripes while experiments of Nuclear
Magnetic Resonance\cite{gorkov} were seen as signatures of
inhomogeneous structures. A low temperature pseudogap detected with
tunneling spectroscopy\cite{naturetunn} points to the presence of a
ground state other than superconductivity whose origin could rely in
inhomogeneous configurations.

In this work we perform a systematic analysis of homogeneous versus
inhomogeneous structures in the electron doped $t-t'$ Hubbard model
using unrestricted Hartree-Fock (UHF) calculations. Stripes and
inhomogeneities were predicted as a solution of the Hubbard model
with $t'=0$ using UHF approximation\cite{uhfstr}. Within this
framework we compute the spectral function and the non-resonant
Raman response of the different configurations. In the study of
commensurate antiferromagnetism  it is interesting to have charge
sensitive probes as the non-resonant Raman response which can detect
charge order due to the possible selection of different regions of
the Brillouin zone\cite{dicastro}. Inhomogeneous solutions in
electron doped have been studied previously with UHF in the three
band Hubbard model\cite{threeband} where they found that diagonal
anti-phase stripes were the configuration lowest in energy. So far
there are no evidence of these stripes in the experiments. We will
use the simplest $t-t'$ Hubbard model which has shown to be able to
explain the pseudogap and the phase diagram of electron
doped\cite{tremblay} cuprates and to give the correct low energy
physics when compared with the two-band Hubbard
model\cite{twobands}. The $t-t'$ Hubbard model has also been studied
with UHF in the past\cite{us,normand} where it was pointed out that
$t'$ favors homogeneous solutions in the electron doped case in
contrast to the hole-doped. This was claimed to be due to the fact
that $t'$ does not frustrate the antiferromagnetism in the electron
doped case as was also proven with other techniques as exact
diagonalization in the electron doped $t-t'-J$
model\cite{tohyama,leung} and in Quantum Monte Carlo calculations of
the $t-t'$ Hubbard model\cite{moreo,huang}. Metallic
antiferromagnetism has been found in experiments\cite{onose}. Our
calculations agree with the former results that homogeneous
antiferromagnetism is more stable in the electron than in hole doped
case but in this systematic study we have found that for {\em large
enough values of the on-site interaction} $U=6t-8t$ and at {\em
intermediate doping} in-phase inhomogeneous configurations have the
lowest energy. This is in strong contrast with what happens in the
hole doped case where inhomogeneities are already found very close
to half-filling\cite{us}. The result is compatible with the recent
theoretical proposal of phase separation at intermediate
fillings\cite{arrigoni}. Phase separation has also been claimed for
the electron doped $t-t'-J$ model\cite{leung94}. Unrestricted
Hartree-Fock can work better to explore the electron doped cuprates
than the hole doped cuprates because when doping the $t-t'$ Hubbard
model with electrons we fill the wider upper antiferromagnetic band
where quantum fluctuations are expected to be less important.

ARPES experiments in electron doped\cite{armitage} cuprates show a
Fermi surface made of electron pockets for  dopings of $x=0.04$,
$x=0.10$ and $x=0.15$ as expected from long range antiferromagnetic
order\cite{moreo,kusko}. They also see the emergence of a hole-like
Fermi surface around the nodal direction at $x=0.15$. This nodal
signal seems to indicate the instability of the
antiferromagnetism\cite{stability,tremblay}. As we work in the
strong coupling limit we do not obtain the hole-like Fermi surface
at $x=0.15$. It is known that to get it in mean-field calculations
unrealistic values of $U$ are needed\cite{kusko}. This case has been
explored with different techniques where quantum fluctuations are
properly taken into account  (see \cite{tremblay,hanke,yuan}) but
these techniques are not well suited for studying inhomogeneities.
On the other hand ARPES also indicates the presence of spectral
weight in the insulator gap. Theoretically, these midgap states have
been obtained by adding spin fluctuations to the mean field
solution\cite{rice}. Inhomogeneous configurations also form midgap
states which would be an alternative explanation of these states.

A related issue concerns recent puzzling experimental results of
electron-doped cuprates. Although the Fermi surface found in ARPES
for most of the electron-doped cuprates  lies in the antinodal
direction, recent Raman experiments have found that coherent
quasi-particles mainly reside in the nodal direction of the
Brillouin zone\cite{ramansc} in all the samples that they have
analyzed ($x=0.135$, $x=0.147$ and overdoped samples). In particular
they see that the $B_{2g}$ signal which involves averages mostly
over the nodal region has a stronger quasiparticle component than
the $B_{1g}$ signal which deals with the antinodal. Both signals
show an incoherent background. In our calculations we find  cases of
inhomogeneous configurations that present a quasiparticle-like
signal in the $B_{2g}$ Raman response although the Fermi level does
not cross the nodal region. We argue that this is an example of a
case in which  two-particle properties cannot be induced from
one-particle properties. We also find that inhomogeneities provide
an incoherent background to the Raman response absent in the case of
homogeneous metallic antiferromagnetism. Our results are
qualitative. We do not pretend to do a close comparison with
experiments for what we would have to take into account other very
important effects as quantum fluctuations, scattering by impurities
etc. We intend to find out the prints of inhomogeneities in the
Raman signal.

The organization of this paper is as follows. In  section II we
review the $t-t'$ Hubbard model and the unrestricted Hartree-Fock
method. In Sect. III we discuss the $U-n$ phase diagram. Sect. IV
gives the results obtained on one-particle properties and Section V deals with
the electronic Raman response. In section VI we present our conclusions and open
problems.

\section{The model and the method}
The $t-t'$ Hubbard model is defined in the two dimensional squared
lattice by the hamiltonian

\begin{equation}
H=-t\sum_{<i,j>s} c^+_{is} c_{js} -t'\sum_{<<i,j>>s} c^+_{is}
c_{js}+U\sum_i n_{i\uparrow} n_{i\downarrow}
 ,
\label{ham}
\end{equation}
which gives rise to  the dispersion relation
$$
\varepsilon({\bf k}) = -2t\;[ \cos(k_x a)+\cos(k_y a)] -4t'\cos(k_x
a)\cos(k_y a) .
$$
 $t>0$ and  $t'=-0.3t$ are generic values for modeling the physics of
the cuprates. We will consider electron doping ($n>1$). We work in
lattices of different sizes  with periodic boundary conditions
depending on the problem at hand. We obtain the $U-n$ phase-diagram
working in a $L=14\times 14$ lattice. We use a $16 \times 16$
lattice to discuss in-phase versus off-phase configurations and a
$L=24\times 24$ lattice to compute the spectral function and the
Raman response. A comparison of the results on the different lattice
sizes for homogeneous and inhomogeneous configurations is done in
the next section. The unrestricted Hartree-Fock approximation
minimizes the expectation value of the hamiltonian (\ref{ham}) in
the space of Slater determinants. These are ground states of a
single particle many-body system in a potential defined by the
electron occupancy of each site. This potential is determined
self-consistently
\begin{eqnarray}
H&=&-\sum_{i,j,s} t_{ij} c^\dag_{is}c_{js} -\sum_{i,s,s'}\frac{U}{2}
{\vec m}_i c^\dag_{is}
{\vec \sigma}_{ss'}  c_{is'} \nonumber\\
&&+\sum_{i}\frac{U}{2} q_i(n_{i\uparrow}+n_{i\downarrow}) +
{\displaystyle c.c.},
\end{eqnarray}
where $t_{i,j}$ denotes next, $t$,  and next-nearest, $t'$,
neighbors, and the self-consistency conditions are
$${\vec m_i}=\sum_{s,s'}<c^\dag_{is} {\vec \sigma}_{ss'}  c_{is'}> ,\;\;\;
q_i=< n_{i\uparrow}+n_{i\downarrow}-1> ,$$ where ${\vec \sigma}$ are
the Pauli matrices.

We will call $E_\lambda$ where $\lambda=1 \ldots 2L$
the eigenvalues of the hamiltonian
and $u_1^\dag|0\rangle,\ldots,u_{2L}^\dag|0\rangle$ the basis that
diagonalizes the hamiltonian. These fermionic operators are obtained
from the original ones through:
\begin{equation}
c_{js}=\sum_\lambda U_{js,\lambda}u_\lambda. \label{eq:chvar}
\end{equation}

\section{$U-n$ phase diagram}
\begin{figure}
\includegraphics[clip,width=0.4\textwidth]{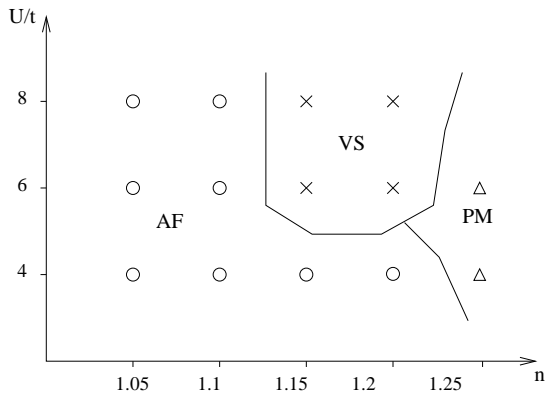}
\caption{$U-n$ phase diagram. The symbols represent the values
studied. AF stands for antiferromagnetism, VS for in-phase vertical
stripes, the inhomogeneous configuration represented in the
Fig.~\ref{fig:conf} (up), and PM for paramagnetism. At strong
coupling and intermediate doping inhomogeneous configurations (VS)
are the dominant ones in the phase diagram.} \label{fig:phUn}
\end{figure}

\begin{figure}
\includegraphics[clip,width=0.4\textwidth]{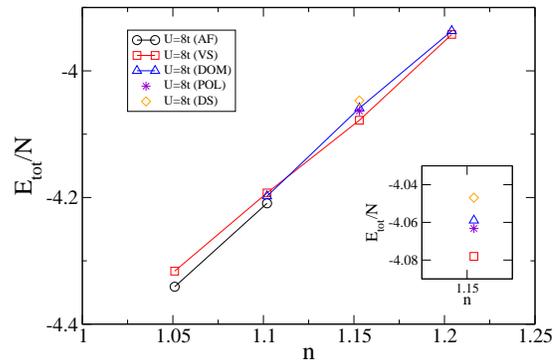}
\caption{(Color online) Comparison of the energy per particle for
different configurations in a $14\times 14$ lattice for different
densities $n$. AF stands for antiferromagnetism, VS for in-phase
vertical stripes (Fig.~\ref{fig:conf} (up)), DOM for in-phase
antiferromagnetic domains (Fig.~\ref{fig:conf} (down)), POL for
polarons and DS for in-phase diagonal stripes. The inset shows the
$U=8t$ and $n=1.15$ region enlarged to appreciate better the
difference in the energies.} \label{fig:compE}
\end{figure}

\begin{figure}
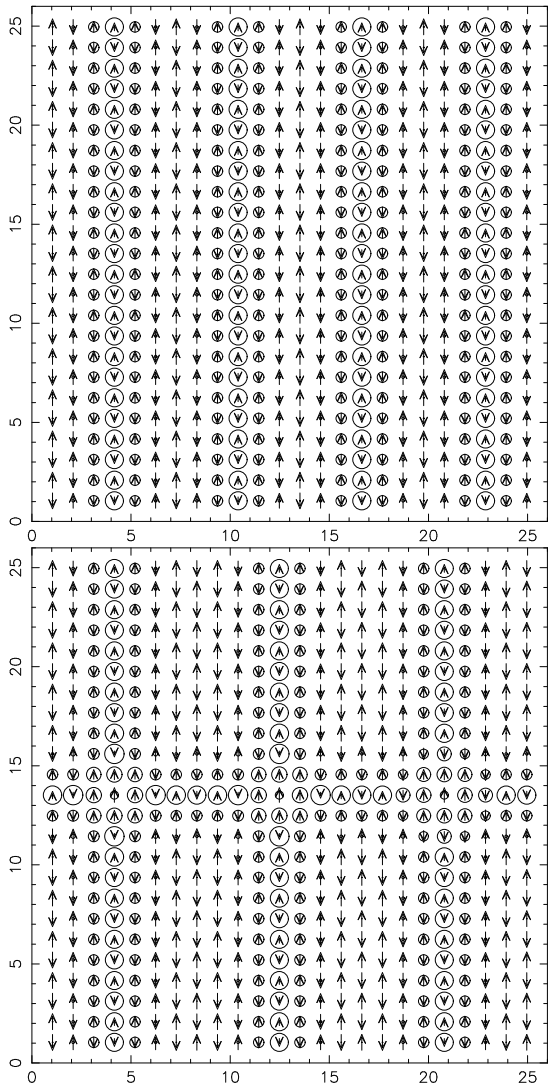

\includegraphics[clip,width=0.4\textwidth,angle=-90]{du8t3vdwe86x24.ps}
\includegraphics[clip,width=0.4\textwidth,angle=-90]{du8t3domce86x24.ps}
\caption{Configuration for $U=8t$, $t'=-0.3t$ and $n=1.15$ in a $24
\times 24$ lattice for in-phase vertical stripes -VS- (up) and
in-phase antiferromagnetic domains -DOM- (down). The circles and
arrows are scaled to the local charge and spin. } \label{fig:conf}
\end{figure}

We have built a detailed $U-n$ phase diagram for $t'=-0.3t$, which
it is shown in Fig.~\ref{fig:phUn} by comparing the energies of
paramagnetic, antiferromagnetic and inhomogeneous configurations in
a $14\times 14$ lattice size. Antiferromagnetism prevails over most
of the phase diagram and inhomogeneities (VS in Fig.~\ref{fig:phUn}
) can be found at intermediate doping and for high values of $U$.
For the inhomogeneous solutions we have compared diagonal (DS) and
vertical (VS) in-phase stripes, in-phase antiferromagnetic domains
(DOM) and polarons (POL). Fig.~\ref{fig:compE} shows the comparison
of the energies per particle for $U=8t$ and $t'=-0.3t$.  All these
inhomogeneous solutions converge for  $n=1.15$ where no
antiferromagnetism can be found. In fact, antiferromagnetism does
not converge in all the VS region in the phase diagram of
Fig.~\ref{fig:phUn}. This is a clear indication that there is a
critical density where inhomogeneities are preferred over
homogeneous antiferromagnetism in unrestricted Hartree-Fock. The
inhomogeneous solution lowest in energy consists of in-phase
vertical stripes (VS) represented in Fig.~\ref{fig:conf} (up) in a
$24 \times 24$ lattice. In-phase antiferromagnetic domains (DOM) are
very close in energy and converge very easily. They are represented
in Fig.~\ref{fig:conf} (down). Magnetic polarons are only stable in
the point $U=8t$ and $n=1.15$. Away from it they are such  that even
if we start to run the system in a polaronic configuration it ends
up in the metallic antiferromagnetism, i.e polarons dilute in the
antiferromagnetic background. This is in contrast with the case for
$t'= 0$ where polarons are the preferred configuration\cite{paco} at
low and intermediate dopings for high values of $U$. The in-phase
diagonal stripe (DS) only converges at $U=8t$, $t'=-0.3t$ and
$n=1.15$ but it is the highest in energy, however DS is the lowest
in energy in hole doped Hubbard model within UHF\cite{us,normand}.
We will further comment on these results when we discuss the density
of states.

To check that the results obtained are robust over the size of the lattice
and as in the following section we will use results for a $24 \times
24$ lattice we have made a comparison of the energy per particle for
different lattice sizes.  The results are shown in
Table~\ref{lattsize}. The compared energy per particle are for
homogeneous antiferromagnetism (AF) and inhomogeneous in-phase
stripes (VS) configurations for two lattice sizes $12\times 12$ and
$24 \times 24$. The lattice size has been chosen for convenience to
fit the inhomogeneous configuration. In the homogeneous case the
variation in energy is almost negligible and in the VS case it is
appreciable in the fourth decimal digit. Therefore we can trust that
the phase diagram will remain for bigger lattice sizes.
\begin{table}
  \centering
  \begin{tabular}{lrc}\hline
  Conf. & $E/L(12\times 12)$ & $E/L(24\times 24)$  \\  \hline
       AF       & -4.46587799  &  -4.46587795 \\ \hline
       VS       & -4.04248799  &  -4.04257066  \\   \hline
  \end{tabular}
  \caption{Energy per site in units of $t$ of different configurations
  for different lattice sizes $12\times 12$ and $24 \times 24$ for $U=8t$ and $t'=-0.3t$.
  In the homogeneous solution (AF) at $n=1$ the difference is negligible and for
  the inhomogeneous solution (VS) at $n=1.16$ it is appreciable in the fourth digit. }\label{lattsize}
\end{table}

Next we compare in-phase with off-phase stripes. We have chosen a
different lattice size ($16 \times 16$) to fit the off-phase
configuration. For the points shown in Fig.~\ref{fig:phUn},
off-phase vertical stripes only converge at $n=1.15$ and $U=8t$ as
an excited state, though the energy per particle is very similar to
the in-phase stripe ($E_{VS}/L=-4.090t$, $E_{STRIPE}/L=-4.083t$).
Therefore, within Unrestricted Hartree Fock, in-phase stripes are
preferred. Adding fluctuations to the unrestricted Hartree-Fock
solution could change this view. Calculations using an exact
diagonalization method within the dynamical mean field
theory\cite{fleck} have shown that solitonic formation contributes
to the stability of the off-phase stripes. However we have noticed a
clear tendency for in-phase solutions that converge easily and in a
wider region of the phase diagram.

\section{one-particle properties}
The spectral function $A({\bf k},\omega)=-\frac{1}{\pi}\Im
G_{ret}({\bf k},\omega)$ is evaluated using the time-dependent Green's
function, which can be calculated numerically.
When the spectral function is expressed in terms of the new
manybody state by Eq.~\ref{eq:chvar} the following expression is
obtained\cite{dagotto}
\begin{equation}
A({\bf k},\omega)=\frac{1}{L}Re \Big[\sum_{{\bf m},{\bf
n},s,\lambda}e^{i({\bf m}-{\bf n}){\bf k}} U_{{\bf n} s,\lambda}
U_{{\bf m}s,\lambda}^\dag\delta(\omega-E_\lambda)\Big].
\end{equation}

\begin{figure}
\includegraphics[clip,width=0.4\textwidth]{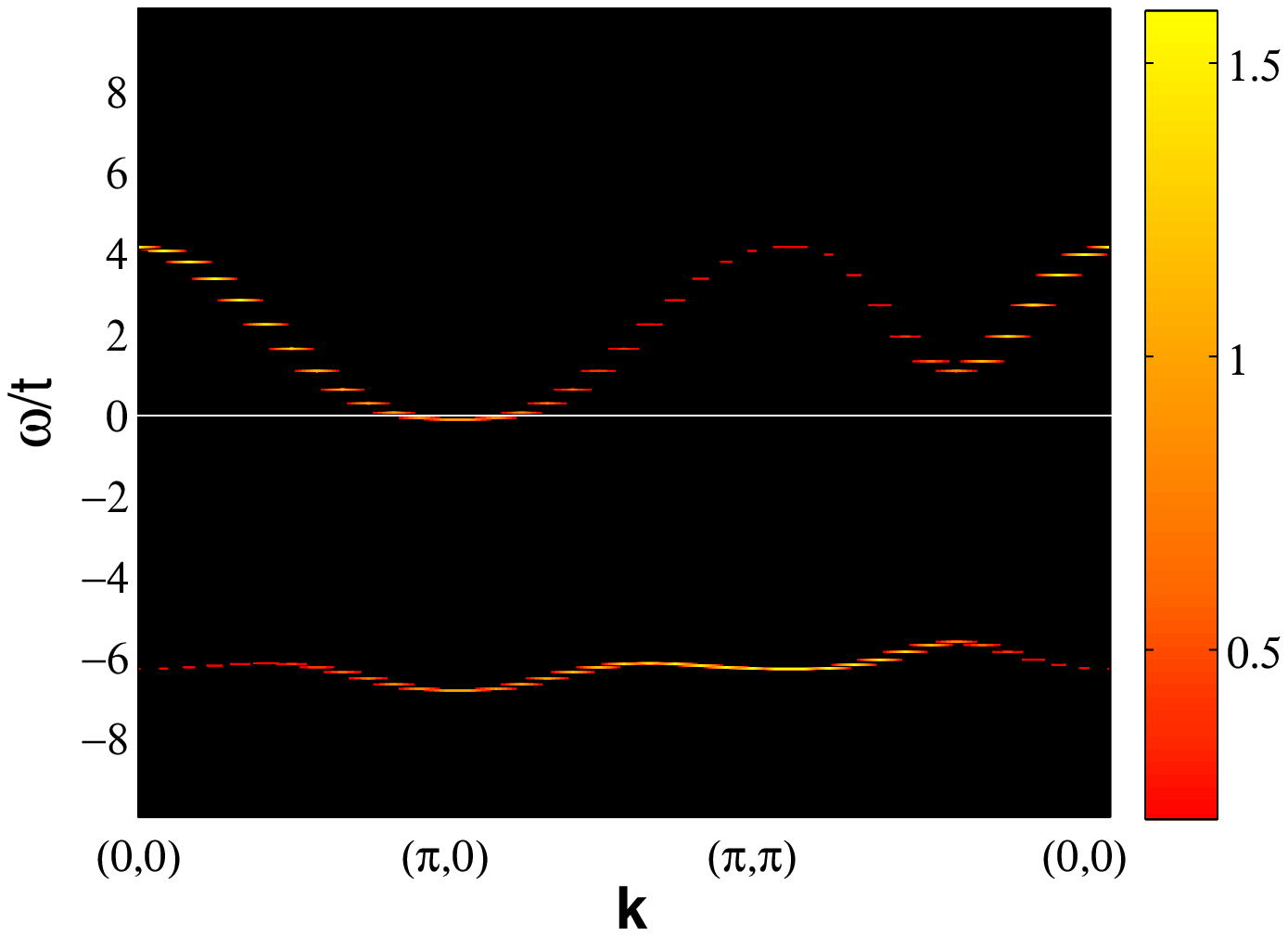}
\includegraphics[clip,width=0.4\textwidth]{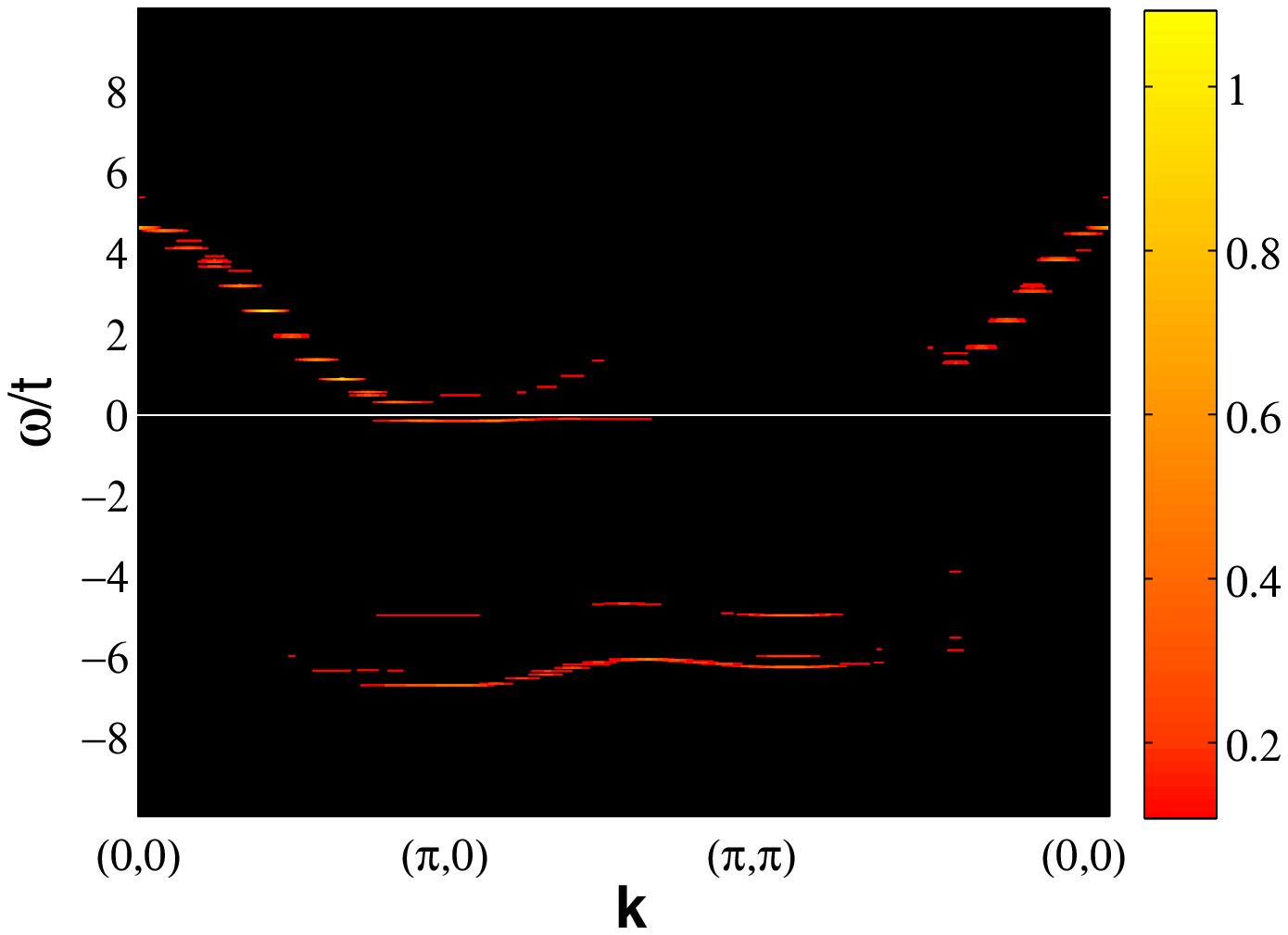}
\caption{(Color online) $A({\bf k},\omega)$ for different
configurations antiferromagnetism at $n=1.05$ (up) and DOM at
$n=1.15$ (down). Parameters are $U=8t$, $t'=-0.3t$ and $T=0.05t$.}
\label{fig:akwvsx}
\end{figure}

We represent $A({\bf k},\omega)$ in Fig.~\ref{fig:akwvsx} with the
Fermi level at $\omega=0$, for AF (Fig.~\ref{fig:akwvsx} up) at
$n=1.05$ and for DOM (Fig.~\ref{fig:akwvsx} down) at $n=1.15$. The
spin-charge configuration of DOM is represented in
Fig.~\ref{fig:conf} (down) in a $24\times 24$ lattice for parameters
$U=8t$ and $t'=-0.3t$. The VS represented in Fig.~\ref{fig:conf}
(up) has a similar spectral density as the DOM. We represent the DOM
spectral function instead the VS because this result will give us a
coherent quasiparticle component in the $B_{2g}$ Raman response
though the Fermi level does not crosses the band in the nodal
region. In Fig.~\ref{fig:akwvsx}(a) we observe two well known
results, (1) $t'$ makes the lower band narrower and the upper band
wider. In consequence, it is expected that localization effects
become more important when we dope with holes than when we dope with
electrons. (2) concerning the ${\bf k}$ dependence for homogeneous
antiferromagnetism, at low doping the Fermi level goes to the bottom
of the upper band at $(\pi,0)$ as suggested in a rigid band
structure. This result is compatible with ARPES\cite{shen}. For
inhomogeneous configurations, Fig.~\ref{fig:akwvsx}(b), we do not
longer have the rigid band structure view and the spectral weight is
reorganized. Midgap states appear and the spectral weight around
$(\pi,\pi)$ in the upper band disappear. The Fermi level is again at
the bottom of the upper band at $(\pi,0)$. The midgap states will be
more appreciated in the density of states.

\begin{figure}
\includegraphics[clip,width=0.4\textwidth]{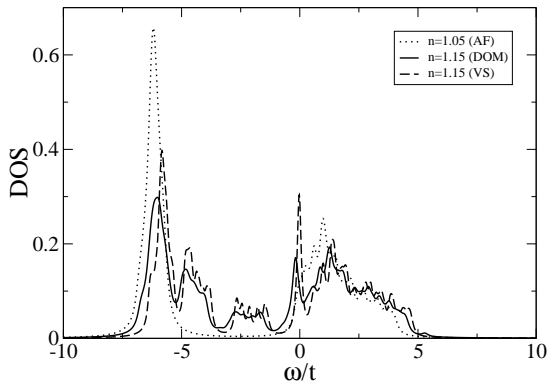}
\caption{Density of states for homogeneous antiferromagnetism at
$n=1.05$ (dotted line), in-phase antiferromagnetic domains DOM
(solid line) and in-phase vertical stripes VS (dashed line) at
$n=1.15$. Both inhomogeneous configurations present similar density
of states with in-gap states and a pseudogap-like feature at the
Fermi level. Parameters are $U=8t$, $t'=-0.3t$, $T=0.05t$ and
$\eta=0.1t$.} \label{fig:dosvsx}
\end{figure}

Fig.~\ref{fig:dosvsx} shows the density of states for the  AF and
DOM configurations described previously and also for the VS
represented in Fig.~\ref{fig:conf} (up) with a width of $\eta=0.1t$
given to the delta functions. The Fermi level is located at zero
energy. We again see for the antiferromagnetic solution (dotted
line) the result expected from a rigid band structure, the gap is
lower if we compare with the result at half filling (not shown) and
the Fermi level is shifted to the upper band. For DOM (solid line)
and VS (dashed line) the density of states is very similar. We
appreciate the formation of midgap states with almost no gap closing
if we compare with the former case at lower doping $n=1.05$. It is
interesting to notice in Fig.~\ref{fig:dosvsx} that for the
inhomogeneous solution there is a depression at the Fermi level (it
can also be seen in Fig.~\ref{fig:akwvsx}(down)). We do not claim
that this corresponds to the pseudogap whose physics needs better
approaches, but it is interesting to realize that inhomogeneous
solutions can give a feature that it is usually considered as a
hallmark of the pseudogap.\cite{tremblay}

Next we discuss further the phase diagram of Fig.~\ref{fig:phUn}
with the help of the spectral function (Fig.~\ref{fig:akwvsx}) and
the density of states (Fig.~\ref{fig:dosvsx}). A simple argument to
understand why the antiferromagnetism is enhanced and dominates the
phase diagram of the electron doped case can be given by changing to a
N\'eels state basis. In that basis the $t'$ term appears in the diagonal
of the Hamiltonian since it represents a hopping within the same
sublattice. As such
it does not change the antiferromagnetic background.
The first electron carrier will go to $(\pi,0)$
(see Fig.~\ref{fig:akwvsx}(up)) with an energy given by $E(\pi,0) =
4t'+\Delta$ where $\Delta=mU$ and $m$ is the magnetization. As
$t'<0$ it lowers the energy of the antiferromagnetic solution.
A similar argument is given in Ref.\cite{tohmaek01}
for the electron doped $t-J$ model.
This simple argument is
reinforced by the study of the stability of the metallic
antiferromagnetic solution\cite{stability}. In the density of states
represented in Fig.~\ref{fig:dosvsx} can be  seen that our
antiferromagnetism is also metallic and the whole picture is very similar
to what happens in a rigid band model.

Next we discuss the VS at
intermediate doping and high $U$. As mentioned when doping with
holes we are doping a narrower band and a lower mobility is expected
than when doping with electrons. It is not surprising then that
inhomogeneities appear at low doping in hole doped and not in
the electron doped. However there is a critical density in the electron
doped case where inhomogeneities are the preferred configuration as a
compromise to minimize the strong Coulomb interaction. In-phase
vertical stripes allow more mobility than diagonal stripes and that
is why within UHF diagonal are preferred for hole
doped\cite{us,normand} and vertical for electron doped.

\section{Non-resonant Raman response }
The differential Raman cross section is\cite{cardona}
\begin{equation}
\frac{\partial^2\sigma}{\partial\omega\partial\Omega}=\frac{\omega_S}{\omega_I}r_0^2
S_{\gamma\gamma}({\bf q},\omega,\beta),
\end{equation}
where $r_0=e^2/mc^2$ is the classical radius of the electron and the
generalized dynamic structure factor is
\begin{equation}
\tilde{S}_{\gamma\gamma}({\bf
q},\omega,\beta)=\sum_{i,f}\frac{e^{-\beta E_i}}{Z}|\langle
f|\tilde{\rho}_\gamma({\bf q})|i\rangle |^2\delta(E_f-E_i+\hbar
\omega). \label{eq:gstrucf}
\end{equation}
Here, $\frac{e^{-\beta E_i}}{Z}$ is the probability of the initial
state in thermodynamic equilibrium. The state $|i\rangle$ is the
initial many-body state and $|f\rangle$ is the final many-body
state. $\tilde{\rho}({\bf q})$ is an effective charge density given
by
\begin{equation}
\tilde{\rho}_\gamma({\bf q})=\sum_{\bf k,s} \gamma({\bf
k};\omega_I,\omega_S)c_s^\dag({\bf k+q})c_s({\bf k}),
\label{eq:effrho}
\end{equation}
where $s$ is the spin index and $\gamma({\bf k};\omega_I,\omega_S)$
is the Raman scattering amplitude. We are interested in the ${\bf
q}=0$ case since the momentum transfer by the photon is negligible.
The generalized structure factor given by Eq.~\ref{eq:gstrucf} can
be interpreted as the power spectrum of the charge density
fluctuations at frequency $\omega$. It is related to the imaginary
part of the Raman response function through the
fluctuation-dissipation theorem
\begin{equation}
S_{\gamma\gamma}({\bf q},\omega)=-\frac{1}{\pi}[1+n(\omega)]{\rm Im}
\chi_{\gamma\gamma}({\bf q},\omega),
\end{equation}
where
\begin{equation}
\chi_{\gamma\gamma}({\bf q},\omega)=\int_0^{1/T}d\tau
e^{-i\omega\tau}\langle T_\tau \hat {\rho}(\tau)\rho(0)\rangle.
\label{eq:chi}
\end{equation}
We will compute the imaginary part of the Raman response by replacing the
Hartree-Fock eigenvalues and eigenvectors in the generalized
structure factor given by Eq.~\ref{eq:gstrucf} instead of using the
Green's function machinery in Eq.~\ref{eq:chi} as it is usually
done.

We will follow Ref.\cite{devereaux} where symmetry is used to
classify the scattering amplitude since the different polarizations
transform as elements of the point group of the
crystal
\begin{equation}
\gamma({\bf k};\omega_I,\omega_S)=\sum_L
\gamma_L(\omega_I,\omega_S)\Phi_L(\bf k),
\end{equation}
where $\Phi_L(\bf k)$ are the Brillouin zone harmonics (BZH). The use of
symmetry arguments instead of the inverse effective mass
approximation as it is usually done\cite{cardona} will allow us
to discuss non-resonant Raman response at high
energies\cite{devereaux}. In this work we are interested in the
$B_{1g}$ and $B_{2g}$ symmetries of the square lattice. The $B_{1g}$
symmetry corresponds to incident and scattered light polarizations
aligned along $\hat{x}+\hat{y}$ and $\hat{x}-\hat{y}$, and thus
\begin{equation}
\Phi_{B_{1g}}(\bf k)=\cos(k_x)-\cos(k_y)+\cdots,
\end{equation}
where the dots stand for higher order BZH. The $B_{2g}$ symmetry
corresponds to incident and scattered light polarizations aligned
along $\hat{x}$ and $\hat{y}$
\begin{equation}
\Phi_{B_{2g}}(\bf k)=\sin(k_x)\sin(k_y)+\cdots,
\end{equation}
The $B_{1g}$ signal weights mainly the antinodal quasiparticles and
the $B_{2g}$ signal the nodal quasiparticles.

Replacing Eq.~\ref{eq:chvar} in Eq.~\ref{eq:effrho} and
Eq.~\ref{eq:gstrucf} we obtain for the $B_{1g}$ and the $B_{2g}$
Raman expressions
\begin{eqnarray}
\lefteqn{{\rm Im}\chi_{B_{1g}}(\omega)=\frac{\pi }{L} \sum_{\lambda
\ne \lambda'} \frac{e^{-\beta E_\lambda}} {(1+e^{-\beta
E_{\lambda'}})(1+e^{-\beta E_\lambda})}} \nonumber\\ & &
\delta(\omega+(E_\lambda-E_{\lambda'}))| \sum_{{\bf
j}=(x,y)s}(U_{(x+1,y) s,\lambda}^\dag
U_{(x,y)s,\lambda'}+h.c)\nonumber\\ & & - (U_{(x,y+1)s,\lambda}^\dag
U_{(x,y)s,\lambda'}+h.c)|^2, \label{eq:b1g}
\end{eqnarray}

\begin{eqnarray}
\lefteqn{{\rm Im}\chi_{B_{2g}}(\omega)=\frac{\pi}{L} \sum_{\lambda
\ne \lambda'} \frac{e^{-\beta E_\lambda}} {(1+e^{-\beta
E_{\lambda'}})(1+e^{-\beta E_\lambda})}}\nonumber\\ & &
\delta(\omega+(E_\lambda-E_{\lambda'}))| \sum_{{\bf
j}=(x,y)s}(U_{(x+1,y-1) s,\lambda}^\dag U_{(x,y)s,\lambda'}+h.c)
\nonumber\\ && -(U_{(x+1,y+1)s,\lambda}^\dag
U_{(x,y)s,\lambda'}+h.c)|^2\nonumber. \label{eq:b2g}
\end{eqnarray}

\begin{figure}[h]
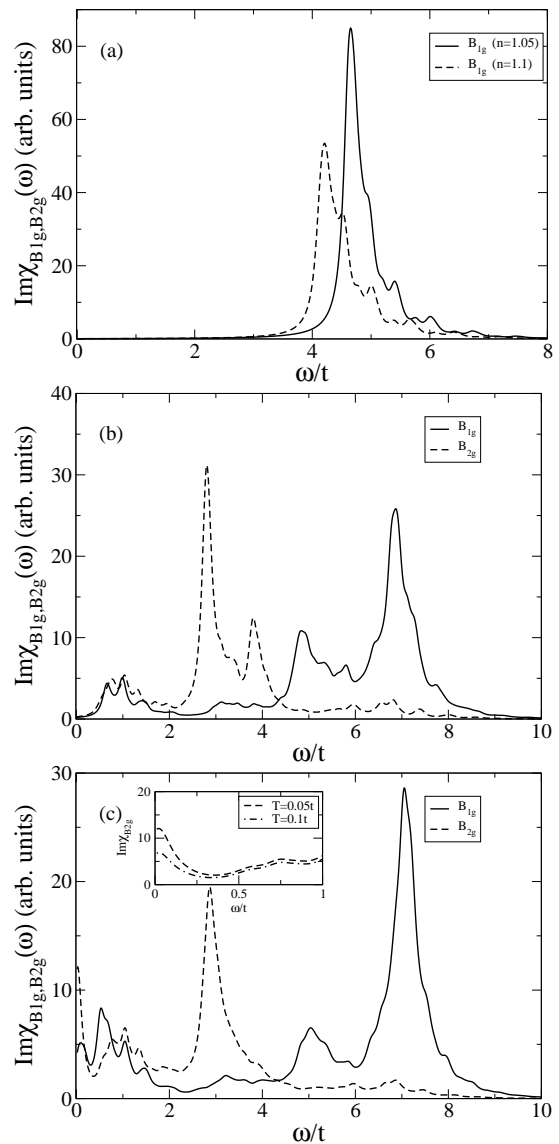

\includegraphics[clip,width=0.4\textwidth]{b1gb2gafx24.eps}
\includegraphics[clip,width=0.4\textwidth]{b1gb2gvdwx24.eps}
\includegraphics[clip,width=0.4\textwidth]{b1gb2gdomx24.eps}
\caption{ (a) Evolution of the $B_{1g}$ and $B_{2g}$ Raman signal
with doping for antiferromagnetism for $U=6t$, $x=0.05$ (solid line)
and $x=0.1$ (dashed line). (b) $B_{1g}$ and $B_{2g}$ Raman signal
for VS (Fig.~\ref{fig:conf} (up)) at $U=8t$ and $n=1.15$. (c)
$B_{1g}$ and $B_{2g}$ Raman signal for DOM (Fig.~\ref{fig:conf}
(down)) at $U=8t$ and $n=1.15$. In the inset is shown the evolution
with temperature of the $B_{2g}$ signal for low frequencies.
Parameters are $t'=-0.3t$, $T=0.05t$ and $\eta=0.1t$. }
\label{fig:raman}
\end{figure}
Fig.~\ref{fig:raman} shows the non-resonant Raman response. In
experiments for cuprates, the range in frequencies is just up to
around $1t$ and at high frequencies the resonant Raman response is
unavoidable and it could couple with the non resonant one as has
been recently shown with Dynamical mean field
theory\cite{devereauxres}, but we show the complete range of
frequencies to illustrate the physics that comes up from our
calculations. We can phenomenologically decompose the Raman response
as following:
$\chi''(\omega)=\chi''_{low\omega}+\chi''_{incoh}+\chi''_{gap}$. The
low frequency response is usually fitted as a quasiparticle response
$\chi''_{QEP}(\omega)=a\frac{\Gamma \omega}{\omega^2+\Gamma^2}$, in
our case $\Gamma = 0$ since we did not put any scattering rate as
input, and therefore $\chi''_{QEP}=0$. But for some cases we find
$\chi_{low\omega} \ne 0$ as we will see in the following.
$\chi''_{gap}$ is formed by the pronounced peak at high energies in
both figures and is due to the antiferromagnetic gap.
$\chi''_{incoh}$ is formed by the rest. In experiments this part is
almost a constant continuum background and gave rise to the Marginal
Fermi liquid phenomenology\cite{mfl}. Next we discuss the results.
Fig.~\ref{fig:raman}(a) represents the evolution with doping of the
$B_{1g}$ and $B_{2g}$ Raman signals for homogeneous
antiferromagnetism with parameters $U=6t$, $t'=-0.3t$ and dopings
$n=1.05$ (solid line) and $n=1.1$ (dashed line). Again a width of
$\eta=0.1t$ is given to the delta functions.  As expected for
homogeneous antiferromagnetism the signal for the $B_{2g}$ channel
is negligible and cannot be appreciated in Fig.~\ref{fig:raman}(a)
and in the $B_{1g}$ channel, $\chi''_{gap}$ is the only appreciable
component. We can observe a shift in $\chi''_{gap}$ for $n=1.1$
compared with $n=1.05$ reflecting the closing of the gap with
increasing doping. That $\chi''_{low\omega}$ and the
$\chi''_{incoh}$ are negligible might be due to the fact that the
charge is completely homogeneous for this antiferromagnetism and
from Eq.~\ref{eq:b1g} and Eq.~\ref{eq:b2g} we see that for
homogeneous phases is very likely that the Raman matrix elements are
zero. We have checked this result with all the homogeneous
configurations that we have found with the same null result for
$\chi''_{low\omega}$ and $\chi''_{incoh}$. Fig.~\ref{fig:raman}(b)
and (c) represents the $B_{1g}$ and $B_{2g}$ Raman signals for the
inhomogeneous solution consisting of VS and DOM at $U=8t$,
$t'=-0.3t$ and $n=1.15$. In both figures a gap is observed at high
frequencies and $\chi''_{incoh}$ is no longer zero neither in
$B_{1g}$ nor in $B_{2g}$ channels. The gap appears at higher energy
in the $B_{1g}$ channel consistent with Fig.~\ref{fig:akwvsx}
(down). The incoherent background comes from the in-gap states in
the density of states, see Fig.~\ref{fig:dosvsx}. For VS there is
not low-frequency component in any channel but for DOM, the $B_{2g}$
channel has a $\chi''_{low\omega}$ component with a
quasiparticle-like shape and it is not clear in the $B_{1g}$
channel. We have calculated the evolution with temperature of the
Raman signal shown in the inset of Fig.~\ref{fig:raman}(c). The
decrease with temperature of the $\chi''_{QEP}$ for the $B_{2g}$
channel would have been interpreted as a metallic behavior in
experiments. We arrive at two interesting results for the
inhomogeneous solution, (1) there is a remarkable incoherent
component, and (2) the $B_{2g}$ channel has a quasiparticle-like
component for the DOM configuration. We can also find this behavior
for higher doping in the VS configuration. This means that the low
energy part of the Raman response can arise not only from the real
quasiparticle component but also  from inhomogeneities. In the last
case this quasiparticle-like response has no relation with real
quasiparticles since the Fermi level does not cross the nodal
region. It might arise because inhomogeneities create an effective
Raman scattering rate that can contribute to the low frequency
region of the Raman response.

\section{Conclusions}
We find in-phase inhomogeneities as the lowest energy configurations
for strong-coupling limit at intermediate dopings. These results are
interesting because the doping region  where inhomogeneities are
found signals  another quantitative asymmetry for hole and electron
doped cuprates. Inhomogeneities are present at very low doping for
hole doped cuprates and only at intermediate to high doping for
electron doped. This region of intermediate doping is precisely the
region where the antiferromagnetic phase is suppressed ($n \approx
1.5$). The result agrees with a recent theoretical proposal of phase
separation for this range of parameters\cite{arrigoni}. It is also
interesting to notice that we need to be in the strong coupling
limit to find the inhomogeneities, since they are formed to minimize
the strong on-site interaction. Experimentally electron doped
cuprates seem to be in the strong coupling limit as much as hole
doped cuprates\cite{millis} but
 there is a theoretical debate on whether or not strong coupling is
necessary to understand electron doped\cite{hanke}. In-phase
inhomogeneities are compatible with experiments of inelastic neutron
scattering and thermal conductivity\cite{ando}. However to fully
address this issue quantum fluctuations should be taken into account
since they can favor off-phase stripes\cite{fleck}.

The inhomogeneities give rise to midgap states in the density of
states as  found by ARPES\cite{armitage} though at strong coupling
it is not possible to obtain a quasiparticle signal in the nodal
region. for what better techniques taking into account quantum
fluctuations should be used. For the non-resonant Raman response the
in-phase inhomogeneities give an incoherent background at low and
intermediate frequencies. For the DOM configuration
(Fig.~\ref{fig:conf} (down)) the $B_{2g}$ signal have stronger
quasiparticle-like component than the $B_{1g}$ signal though the
Fermi level does not cross the nodal region. This is an example
where two particle properties can not be inferred from one particle
properties. It would be interesting to check if the quasiparticle
component of the $B_{2g}$ Raman response is different from zero in
the region where ARPES does not show quasiparticle weight in the
nodal region.

\acknowledgments

We wish to thank M. Moraghebi, T. Devereaux, P. L\'opez-Sancho, F.
Guinea for valuable discussions and to M.A.H. Vozmediano and E.
Bascones for a critical reading of the manuscript. Funding from MCyT
(Spain) through grant MAT2002-0495-C02-01 is acknowledged.


\begin{thebibliography}{99}
\bibitem{dagotto} The development of the Green
function can be followed in detail in the book by Dagotto: Nanoscale
phase separation and colossal magnetoresistance : the physics of
manganites and related compounds, chapter 7. Section 7.5. I have
applied the same ideas to find the Raman response with the
unrestricted Hartree Fock solutions.
\end{thebibliography}
\end{document}